\documentclass[useAMS,usenatbib]{mn2e}
\usepackage{amssymb,graphicx,amsmath}

\newcommand{\qm}[1]
{``#1''}

\title[Modelling Jet Power in Elliptical Galaxies]{Models for jet power in elliptical galaxies: A case for rapidly spinning black holes}

\author[Nemmen et al.]{R. S. Nemmen,$^{1}$\thanks{E-mail: rodrigo.nemmen@ufrgs.br} R. G. Bower,$^{2}$ A. Babul,$^{3}$ and T. Storchi-Bergmann$^{1}$ \\
$^{1}$Instituto de F\'isica, Universidade Federal do Rio Grande do Sul, Campus do Vale, Porto Alegre, RS, Brazil \\
$^{2}$Department of Physics, Institute for Computational Cosmology, Durham University, South Road, Durham DH1 3LE, England \\
$^{3}$Department of Physics \& Astronomy, University of Victoria, Elliott Building, 3800 Finnerty Road, Victoria, BC V8P 1A1, Canada}

\begin{document}

\date{Accepted 2007 March 12. Received 2007 February 24; in original form 2006 August 12}

\pagerange{\pageref{firstpage}--\pageref{lastpage}} \pubyear{2006}

\maketitle

\label{firstpage}

\begin{abstract}

The power of jets from black holes are expected to depend on both the
spin of the black hole and the structure of the accretion disk in the
region of the last stable orbit. We investigate these dependencies
using two different physical models for the jet power: the
classical Blandford-Znajek (BZ) model and a hybrid model developed by
Meier.  In the BZ case, the jets are powered by magnetic
fields directly threading the spinning black hole while in the hybrid
model, the jet energy is extracted from both the accretion disk as
well as the black hole via magnetic fields anchored to the accretion
flow inside and outside the hole's ergosphere. The hybrid model takes
advantage of the strengths of both the Blandford-Payne and BZ
mechanisms, while avoiding the more controversial features of the
latter. We develop these models more fully to account for general
relativistic effects and to focus on advection-dominated accretion flows (ADAF) for
which the jet power is expected to be a significant fraction of the
accreted rest mass energy.

We apply the models to elliptical galaxies, in order to see if these
models can explain the observed correlation between the Bondi accretion
rates and the total jet powers. For typical values of the disk
viscosity parameter $\alpha \sim 0.04 - 0.3$ and mass accretion rates
consistent with ADAF model expectations, we find that the observed
correlation requires $j \gtrsim 0.9$; i.e., it implies that the black holes
are rapidly spinning.  Our results suggest that the central black
holes in the cores of clusters of galaxies must be rapidly rotating in
order to drive jets powerful enough to heat the intracluster medium
and quench cooling flows.

\end{abstract}


\section{Introduction}

Recent high resolution \textit{Chandra} observations of the cores of elliptical galaxies provide dramatic illustration of the impact of jets launched from active galactic nuclei (AGN) on the interstellar medium of their host galaxies and/or on the intracluster medium if the galaxies reside in clusters.  These observations have revealed prominent X-ray surface brightness depressions corresponding to cavities or bubbles inflated by the jets as they interact with the surrounding hot gas (e.g., \citealt{allen06}, hereafter A06; \citealt{birzan04,fabian06,taylor06,rafferty06}).  This interaction is presumed to deposit large amounts of energy in their environments (e.g., \citealt{churazov02,dv04,sijacki06, nusser06}), modifying the hierarchy of galaxy formation  (e.g., \citealt{bower06}) and altering the evolution of the intracluster medium by counteracting radiative losses (e.g., \citealt{mccarthy04}).   Based on the energetics of creating the observed cavities, the minimum energy associated with the jets range from $\sim 10^{55}$ ergs in galaxies, groups, and poor clusters to $\sim 10^{60}$ ergs in rich clusters, ranking the outbursts as  among the most powerful phenomena in the Universe

While a detailed understanding of the extragalactic AGN jet phenomena remains elusive, the combination of physically insightful analytic studies (e.g., \citealt{bz77, bp82, begelman84, punsly90, ferrari98, meier99, meier01}) and sophisticated general relativistic, magnetohydrodynamic (MHD) numerical simulations (e.g., \citealt{koide00, koide03,mckinney04,devilliers05,komissarov05,hawley06}) are beginning to yield important insights. There is now a general consensus that jets are fundamentally MHD events. 

The currently favoured models presuppose an accretion flow threaded by large-scale magnetic fields flowing onto a supermassive black hole.   In the neighborhood of the black hole, the flow settles into a disk-like structure, in which the rate of inward flow of matter depends on the efficiency with which its angular momentum can be transferred outward.  This transfer of angular momentum is mediated by MHD turbulence.  Both MHD turbulence and differential rotation of the plasma in the body of the disk generate and intensify toroidal magnetic fields \citep{balbus98}.   When the pressure associated with the toroidal fields grows strong enough, the field lines escape from the disk forming a rotating helical tower of field lines above and below the disk.  Centrifugal forces associated with this rotating magnetic field helix drives any plasma trapped onto the field lines upward and out of the disk, generating outflows.   This is the crux of the Blandford-Payne \citep{bp82} model for jets, wherein the magnetic fields extract energy from the rotation of accretion disk itself to power the outflows.  Depending on the detailed structure of the magnetic fields and the accretion disk, this mechanism is expected to generate outflows ranging from broad, uncollimated winds to highly collimated jets.  

In the event that the black hole itself is spinning, the magnetic field can also extract the rotational energy of the central black hole to power the outflows.  In the Blandford-Znajek (BZ) model \citep{bz77}, the magnetic fields are assumed to be connected directly to the horizon of a spinning black hole while in the  \citet{punsly90} model, the magnetic fields associated with jet production are anchored to the inflowing plasma inside the black hole's ergosphere.  In both these models, the dragging of inertial frames, relative to an observer at infinity, within the ergosphere of  the rotating black hole results in a rotating, tightly wound vertical tower of field lines, and hence powerful outflows.  Moreover, \citet{meier99} has shown that in the Punsly-Coroniti-like models, the differential dragging of the frames will also act as a dynamo to amplify the magnetic field at the expense of the black hole's rotational energy and that this will have the effect of further enhancing the jet power. The results of numerical simulation studies  \citep{koide03,mckinney04,devilliers05,hawley06} are, in a broad brush sense, consistent with the BZ and Punsly-Coroniti-Meier type of models, indicating that the power of jets depends on both the mass accretion rate as well as the spin of the black hole. 

In fact, \citet{meier01} argues that understanding the radio-loud and radio-quiet dichotomy in the QSO population requires the jet power to exhibit such dependencies.   And while this assertion appears to be indicated by observations of black hole candidate systems in our Galaxy \citep{cui98}, very little  is known about the prevailing conditions underlying the jet phenomena in extragalactic  AGNs.  The recent \textit{Chandra} observations of bubbles, however, offer a unique opportunity to remedy this.   The X-ray observations not only provide an estimate of the power of the jets emitted by the AGNs but also the rate at which matter is accreting onto the black hole-accretion flow system.   Recently \citet{allen06} (hereafter A06) analyzed the data for nine nearby, X-ray luminous elliptical galaxies and found a remarkably tight correlation between the Bondi accretion rates and the jet powers.   Within the context of viable models for the jet-black hole-accretion flow system, such a correlation not only provides insight into the efficiency with which the rest energy of the material accreting onto the black hole is converted into jet energy but also the spin distribution of the black holes powering AGN.  In this paper, we seek to shed light on these issues. 

We begin by considering two physical models for the jet power: the BZ model as described above, and a hybrid model proposed by \citet{meier01} (see also, \citealt{punsly90,meier99}), in which the magnetic field threads the plasma throughout --- in the body of the disk where frame-dragging is negligible as well as within the ergosphere.   This latter model combines the Blandford-Payne-like disk acceleration mechanism with the Blandford-Znajek-like scheme that draws upon the rotational energy of the black hole. We improve upon these models by incorporating important general relativistic effects that previously were introduced via highly restricted approximation.  Following \citet{meier01}, we further couple our models for the jet production to the  advection-dominated accretion flow model (hereafter ADAF \footnote{Improved models of advection-dominated accretion flows incorporating winds and convection are also more generally referred to as radiatively inefficient accretion flows (RIAF). We simply use the original acronym ADAF.}, \citealt{narayan05,narayan98}, hereafter N98; \citealt{nemmen06}).   There is a considerable body of work indicating that the launching of jets is most efficient when the accretion flow is advection-dominated (e.g., \citealt{rees82,meier01,churazov05}) and that jet production is suppressed in the standard thin accretion disk normally associated with radiatively efficient AGNs  \citep{livio99,meier01,maccarone03}.

This paper is organized as follows: In  \textsection \ref{sec:model}, we describe our models for the accretion flow and the jet power.  We summarize the findings of A06 in \textsection \ref{sec:correl}.   In \textsection \ref{sec:constr}, we compare the correlations between the jet power and accretion rate implied by our models with the results of A06 and consider the resulting implications for jet efficiencies and the distribution of black hole spins.  We also compare results for accretion models with and without winds and explore the implications of uncertainties in A06 estimates of jet power and mass accretion rates on our findings. Lastly, we highlight our main results in \ref{sec:conclusion}  and offer some concluding remarks.

\section{Models of accretion flow and jet power} \label{sec:model}

\subsection{Accretion flow structure} \label{sec:riaf}

There are several lines of evidence that suggest that the accretion
flow onto AGNs with strong jets is best described as an ADAF: First,
the observed bolometric (radiative) luminosities of radio-loud AGNs
are typically many orders of magnitude smaller than the luminosities
expected if the mass were flowing onto the central black holes along
thin accretion disks (e.g., \citealt{dmt03,taylor06}); 
in the latter case, the
bolometric luminosities correspond to roughly 10\% of the rest-mass
energy of the accreting matter \citep{shakura73}.
ADAFs can not only account for the observed radiative quiescence of
the jet-emitting AGNs but also reproduce their nuclear X-ray
luminosities with accretion disk mass flow rates\footnote{We use the term
`mass flow rate' to distinguish the mass flowing through the 
accretion disk from the mass that is accreted onto the black hole
itself. Due to the mass/energy carried away by winds and jets,
these two rates may differ significantly.} comparable
to their Bondi rates \citep{dmt01b,dmt03,loewen01,p05,soria06}.
The exceptional systems (such as the radio-loud quasars) that are 
both highly luminous and radio-loud are currently an intriguing puzzle: possibly they are high accretion rate systems in which the disk is puffed up by the trapped energy despite the system's low opacity (e.g., \citealt{meier01,maccarone03,kording06,sikora07}).

Second, the magnitude and the structure of the magnetic fields
associated with ADAFs are much more conductive to the extraction of
spin energy from the hole than those associated with standard thin
disks (e.g., \citealt{rees82,livio99,armitage99}).  Third, X-ray
binaries with black hole candidates (XRBs) in the \qm{low/hard} state
display a strong correlation between the X-ray and radio emissions,
which can best be understood in the context of an ADAF-jet system
\citep{gallo03,merloni03,falcke04}.  Low-luminosity AGNs with jets show
similar correlations and hence, are thought to be scaled-up versions
of the galactic XRBs \citep{merloni03,falcke04}.  One characteristic
of the ADAF-jet system is low mass flow rates.  The observations
suggest that at high flow rates, the black hole systems switch to
the \qm{high/soft} state and the jet activity is suppressed
\citep{maccarone03,greene06}.  This observed \qm{quenching} of the jet
(as well as the transition from the \qm{low/hard} to the
\qm{high/soft}) is believed to be due to a change from an ADAF
(powerful jets) at low mass flow rates to a thin accretion disk
structure (weak jets) at high mass flow rates (e.g., \citealt{meier01}).

We describe the structure of the ADAF using the self-similar equations of \citet{narayan95}.   We are especially interested in the analytical equations (see Appendix I) that describe the vertical half-thickness of the disk $H$, the angular velocity of the disk $\Omega'$ and the magnetic field strength $B$ near the black hole in terms of radius $R$, black hole mass $M_\bullet$, accretion rate onto the black hole $\dot{M}$ and advection parameter $f$ (assumed $\approx 1$).  We require these as inputs to our models for the jet power described in \textsection \ref{sec:bz} and \ref{sec:disk} below.   We note that although the \citet{narayan95} equations are based on an early model for the ADAF that did not allow for mass loss in the form of winds from the accretion flow,  we have verified that these self-similar solutions provide a good approximation to the structure of the inner regions of ADAF models, like that of \citet{bb99} (hereafter BB99), that do allow for such losses.

The equations for $H$, $B$ and $\Omega'$ also depend on the properties of the accreting fluid, such as its adiabatic index $\gamma$, its viscosity parameter $\alpha$ and the ratio of gas to magnetic pressure $\beta$ in the fluid.  These quantities are not independent.  The value of $\gamma$ depends on $\beta$ via the relationship $\gamma=(5 \beta + 8)/3(2+\beta)$ \citep{esin97} and based on the MHD numerical simulations of the evolution of the magnetorotational instability in accretion disks \citep{hawley95}, $\alpha \approx 0.55/ (1+\beta)$.   We assume that the magnetic pressure is related to the field strength as $P_{\rm mag} = B^2/8 \pi$.

We constrain the value of $\alpha$ from recent MHD numerical simulations of radiatively inefficient accretion flows, which take into account self-consistently the role of the Maxwell stresses in establishing its value. The general relativistic simulations (e.g., \citealt{mckinney04,hirose04,hawley06}) find that in the inner portions of the disk $\beta \approx 1-10$. Although $\alpha$ is not explicitly quoted, we can estimate $\alpha$ from these values of $\beta$ using the relationship between the two, obtaining $\alpha \approx 0.04 - 0.3$.   For computational purposes, we will derive results for $\alpha=0.04$ and $0.3$.   To put these $\alpha$-values in context, we note that the simulations around Schwarzschild holes of \citet{proga03} suggest that near the innermost stable circular orbit, $\alpha$ reaches high values: $\alpha \approx 0.1-0.7$. Moreover, recent ADAF models of XRBs 
require values of $\alpha \approx 0.25$ in order to account for the observations \citep{quataert99}.

To the above model for the ADAF, we introduce three important modifications to take into account general relativistic effects induced by the Kerr metric:  First, we take as input for our jet model the values of $H$, $B$ and $\Omega'$ evaluated at $R_{\rm ms}$, the radius of the marginally stable orbit of the accretion disk.  This radius depends sensitively on the dimensionless black hole spin parameter $j \equiv J/J_{\rm max}$  (\qm{$a/M$} in geometrized units, $a$ is the specific angular momentum), where $J$ is the angular momentum of the hole and $J_{\rm max}=GM_\bullet^2/c$ is the maximal angular momentum \citep{bardeen72}.  Second, an observer at infinity will see the disk and the magnetic fields near the black hole rotate, in the Boyer-Lindquist coordinate system, not with an angular velocity $\Omega'$ but $\Omega = \Omega' + \omega$, where
$\omega \equiv  -g_{\phi t} / g_{\phi \phi}$ is the angular velocity, in the same coordinate system, corresponding to the  local spacetime rotation enforced by the spinning black hole \citep{bardeen72}.
And third, we take into account the field-enhancing shear caused by frame-dragging when calculating the magnetic field strength in the inner region of the disk, as first suggested by  \citet{meier99} and tentatively observed in recent MHD numerical simulations \citep{hawley06}.  Following \citet{meier01}, we relate the amplified, azimuthal component of the magnetic field to the unamplified magnetic field strength derived from the self-similar ADAF solution as $B_\phi = g B$, where $g=\Omega/\Omega'$ is the field-enhancing factor.  The amplitude of this factor depends on the black hole spin through the angular velocity of spacetime rotation, $\omega$.   In the case of a non-rotating black hole, $\omega=0$ and $g=1$ (i.e., no field enhancement).  

The potential importance of these effects has been recognized previously and \citet{meier01} even incorporated them into his jet power model,  albeit in the form of simplifications that blunted their impact. For example, he took $g$ to be a free parameter rather than explicitly relate it to spacetime rotation induced by the rotating black hole, he adopted a simple approximation for $\omega$, and he evaluated his jet model only for two limiting radii:   $7GM_\bullet/c^2$ (corresponding to $j \approx 0$) and $1.5GM_\bullet/c^2$ ($j \approx 1$).   In our approach, we are able to capture more fully the dependence of the important model parameters on $j$ and by doing so, are able to explore the sensitivity of the accretion flow and the jet power solutions to variations in the value of the black hole spin parameter.

Finally, we follow \citet{livio99} in assuming that the poloidal and azimuthal components of the magnetic field are related to each other as $B_p \approx H/R \; B_\phi$ where $R$ is the radius in cylindrical coordinates, a relationship based on the assertion that the strength of the poloidal field is limited by the vertical extent of turbulent eddies in the disk.   In the case of an ADAF, this yields $B_p \approx B_\phi$ because $H \sim R$.
 
\subsection{The Blandford-Znajek jet model} \label{sec:bz}

According to the BZ model, the total power of the resulting jet is given by (e.g., \citealt{macdonald82,tpm86}):
\begin{equation} \label{eq:powerbz}
P_{\rm jet}^{\rm BZ} = \frac{1}{32} \omega_F^2 B_\perp^2 R_H^2 j^2 c, 
\end{equation} 
where $R_H =[ 1+(1-j^2)^{1/2} ] G M_\bullet/c^2$ is the horizon radius, $B_\perp$ is the strength of the magnetic field normal to the horizon, and the factor $\omega_F \equiv \Omega_F (\Omega_H-\Omega_F)/\Omega_H^2$ depends on the angular velocity of the field lines $\Omega_F$ relative to that of the hole, $\Omega_H$.    Following the usual practice, we assume that $\omega_F = 1/2$, which maximizes the power output (e.g., \citealt{macdonald82,tpm86}).  As for $B_\perp$, \citet{livio99} has argued that the field threading the horizon ought to be  comparable in strength to the field threading the inner regions of the accretion flow.  Hence, we take the field strength at the horizon to be the same as that at the radius of the marginally stable orbit of the accretion disk; that is, $B_\perp \approx B_p (R_{\rm ms}) \approx g(R_{\rm ms}) B(R_{\rm ms})$. These field configurations are consistent with those seen in the in the jet launch regions in numerical simulations of accretion onto Kerr black holes (e.g., \citealt{hirose04,mckinney04}). Finally since the accretion rate enters the jet power only through the field strength, the jet power only depends on the accretion rate measured at the marginally stable orbit of the accretion flow $\dot{M}_{\rm ms} \equiv \dot{M}(R_{\rm ms})$.  If the mass is conserved in the disk as in the early version of the ADAF model (\citealt{narayan95}, N98) then $\dot{M}$ is constant with radius, but contemporary ADAF models often allow for mass loss from the accretion disk in the form of winds and outflows (see \textsection \ref{sec:power}).

The primary advantage of the BZ model is its simplicity.   The picture describes a black hole whose event horizon is equivalent to a rotating conducting surface with surface charges and currents.   In effect, the horizon is a unipolar inductor onto which the magnetic fields are attached.  The magnetic fields are torqued by the rotation of the black hole and in turn, drive an outflow.   There are, however, concerns about the model.   \citet{punsly90} have argued that the very notion of flux being emitted from the event horizon renders the model unphysical because the event horizon is casually disconnected from events upstream.   Disturbances generated at the event horizon can only propagate inward.   Moreover, the BZ model is also formulated on the assumption of zero accretion that underlies the BZ model; the BZ model is derived in the limit of vanishing plasma density and pressure \citep{hawley06}.   On the other hand, \citet{mckinney04} have reported good agreement between several aspects of the BZ model and their numerical simulation results, and for this reason, we consider this model in the present work.   We do note, however, that \citet{mckinney04} also reported a growing difference between the model and their results with increasing black hole spins, a result that can partly be attributed to the fact that the BZ model is derived in the limit of slowly rotating black holes.

\subsection{Hybrid model} \label{sec:disk}

Contrary to the premise of very low plasma densities that underlies the BZ model, accretion of matter is presumed to be a key element of real AGN systems.   In fact, numerical simulations show that the coupling between the accretion flow and the magnetic fields is an essential element of jet production.   For this reason, we consider a second jet model, the hybrid model of \citet{meier01}.    As noted previously, this model is constructed so that large-scale magnetic fields thread the accretion disk outside the ergosphere as well as rotating plasma within the ergosphere that is flowing onto the black hole.   Hence, the model is able to draw upon both the rotational energy of the accretion disk as well as the spinning black hole in order to drive outflows, though the extraction of the black hole rotational energy occurs indirectly through field lines anchored to the plasma subject to spacetime rotation imposed by the black hole.  
This model also takes into account the effects of field amplification by both the differential rotation of the plasma in the body of the disk and the differential frame-dragging.  As discussed by \citet{meier01}, the jet power in this model is a strong function of the thickness of the accretion disk and the black hole spin; strong magnetic fields and rapid rotation, the necessary ingredients for the launching of powerful jets, only arise when the disk is thick and the hole is spinning rapidly.

Following \citet{meier01}, the total  jet power for the hybrid model is given by 
\begin{equation} \label{eq:powerdisk}
P_{\rm jet}^{\rm disk} = \left( B_{\phi} H R \Omega \right)^2 / 32c,
\end{equation} 
where $B_\phi = g B$ and  $\Omega = \Omega' + \omega$. All quantities are evaluated at $R=R_{\rm ms}$, which is also assumed to be the approximate characteristic size of the jet-formation region. We note that we denote the jet power for this model with the superscript \qm{disk} in order to highlight that in addition to drawing upon the rotational energy of the spinning black hole, this model also draws energy from the accretion disk.

As a matter of interest, we note that there is considerable support for the Punsly-Coroniti-Meier component of the hybrid model from recent numerical simulations studies \citep{koide03,mckinney04,devilliers05}.  On the other hand, the  relevance of the Blandford-Payne mechanism, which describes the extraction of jet power from the rotation of the accretion disk and which is incorporated in our hybrid model, is very much a matter of debate.   For example, \citet{hawley06} have recently argued that this mechanism is not at all important for understanding the collimated jet outflows seen in their numerical simulations while \citet{blandford05} has suggested that such conclusions are premature and that the simulation results are dependent on the initial conditions adopted.

\subsection{Jet models properties} \label{sec:power}

There are several interesting properties of our jet models that ought to be highlighted.  First, the jet power for neither the BZ nor the hybrid models (Equations \ref{eq:powerbz}  and \ref{eq:powerdisk}, respectively) 
depends on the black hole mass.  Moreover, the dependence on the accretion rate enters the jet power only through the field strength (see Appendix I) and for the combined accretion flow-jet power model under consideration here, the jet powers for both models can be expressed as $P_{\rm jet} (\alpha, j, \dot{M}_{\rm ms}) \propto \dot{M}_{\rm ms}$; that is, the jet power depends linearly on the mass accretion rate.  However, both have a complicated dependence on $j$ and $\alpha$ (see Appendix I).  Figure \ref{fig:newpower} illustrates the spin dependence of both the BZ and the hybrid $P_{\rm jet}$ models, for viscosity parameters $\alpha=0.04$ and $0.3$, and a fiducial value of $\dot{M}_{\rm ms}=10^{-3} \; M_\odot \, {\rm yr}^{-1}$. This corresponds to one tenth of the mean Bondi accretion rate of the elliptical galaxies studied by A06.

\begin{figure}
\centering
\includegraphics[scale=0.7]{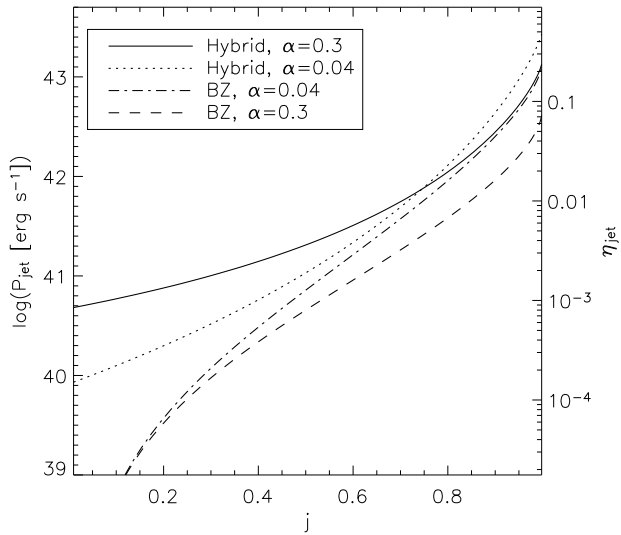}
\caption{The jet powers for our modified Blandford-Znajek (Equation \ref{eq:powerbz})  and the hydrid model (Equation \ref{eq:powerdisk}) as a function of black hole spin parameter $j$.   The curves are computed assuming a fiducial value of $\dot{M}_{\rm ms}=10^{-3} \; M_\odot \, {\rm yr}^{-1}$, and for two different values of the viscosity parameter: $\alpha=0.04$ and $0.3$.  The right axis shows the jet efficiency $\eta_{\rm jet} \equiv P_{\rm jet}/\dot{M}_{\rm ms} c^2$.   Since the jet powers scale linearly with $\dot{M}_{\rm ms}$, $\eta_{\rm jet}$ depends only on $j$ and $\alpha$.}
\label{fig:newpower}
\end{figure}

As Figure \ref{fig:newpower} shows, both the BZ and the hybrid $P_{\rm jet}$ models behave similarly and give comparable results for intermediate and high values of $j$.  The slight difference in the jet power between the two models is not significant given the uncertainty in our understanding of the detailed mechanisms underlying jet formation.   At low spin values, however, the jet power in the hybrid model is considerably higher than that for the BZ case.  The drop-off in the BZ case reflects the decline in the black hole rotational energy available to power the jet while in the hybrid case, the decline is limited by contributions from the accretion disk.     The most important point made by Figure \ref{fig:newpower} is that the jet power for both the BZ and the hybrid models is a strong function of the black hole spin, spanning a range of three orders of magnitude as the black hole spin varies from 0 to 1. This strong dependence is largely due to our improvements over previous models, which did not incorporate carefully the physics of the Kerr metric \citep{ghosh97,armitage99,meier01}.  This type of $j$-dependence for the jet power has previously only been seen in complex numerical MHD simulations of jet formation \citep{mckinney05,hawley06}.

The right axis of Figure \ref{fig:newpower} shows the jet efficiency
factor, defined as $\eta_{\rm jet} \equiv P_{\rm jet}/\dot{M}_{\rm ms}
c^2$.  Since our model jet powers scale linearly with $\dot{M}_{\rm
ms}$, the corresponding jet efficiency factor depends only on the black
hole spin parameter $j$ and the viscosity parameter $\alpha$.  For
high values of $j$, the efficiencies predicted by our two jet models
are comparable to the Novikov-Thorne thin disk radiative efficiency,
which is related to the binding energy of the innermost stable
circular orbit \citep{novikov73}, and the jet carries away a significant
fraction of the rest mass energy flowing through the disc.
We note that for the maximal black
hole spin, $j=0.998$, as implied by the analysis of \citet{thorne74},
the jet efficiencies are $\eta=0.22$ (BZ model) and $0.48$ (hybrid
model) for $\alpha=0.04$.  On the other hand, if $\alpha=0.3$, the
efficiencies drop to $\eta=0.07$ (BZ model) and $0.24$ (hybrid model).

\section[]{The empirical $\dot{M}$--$P_{\rm jet}$ correlation} \label{sec:correl}

Given a distribution of gas about a central black hole,  the most simple configuration describing the accretion of the gas onto the the black hole is the Bondi flow model \citep{bondi52}, which assumes a non-luminous central source and a spherically symmetric flow with negligible angular momentum.    The resulting Bondi accretion rate can be written as $\dot{M}_{\rm Bondi}=\pi \lambda c_s \rho r^2_A$, where $r_A=2GM_\bullet/c_s^2$ is the accretion radius, $G$ is the gravitational constant, $M_\bullet$ is the black hole mass, $c_s$ is the sound speed of the gas at $r_A$, $\rho$ is the density of gas at $r_A$ and $\lambda$ is a numerical coefficient that depends on the adiabatic index of the gas. This estimate is frequently used in studies of the central X-ray emitting gas in elliptical galaxies (e.g., \citealt{dmt03,p05}).

Recently, A06 derived the Bondi accretion rates and jet powers from
\textit{Chandra} X-ray observations of nine nearby, X-ray luminous
giant elliptical galaxies that show evidence of jet-inflated cavities
in their central regions, and found a tight correlation between the
two.  The jet powers were estimated from the energies and timescales
required to inflate cavities observed in the surrounding X-ray
emitting gas such that $P_{\rm jet} = E/t_{\rm age}$, where $E$ is the
energy required to create the observed bubbles and $t_{\rm age}$ is
the age of the bubble.  A06 estimates are based on the assumption that
the X-ray bubbles are inflated slowly.  The values of $\dot{M}_{\rm
Bondi}$ for the systems were calculated from the observed gas
temperature and density profiles. In most cases, the bondi radius is
not observed directly, and the appropriate density and temperature are
determined by extrapolating the observed data, typically by a factor of 3
or greater, in radius (cf., \citealt{rafferty06}). The black hole
masses were deduced from the optical velocity dispersion measurements using
the correlation between central black hole mass and velocity
dispersion of \citet{tremaine02}. Since A06 presents a state of the art 
analysis of current data, we accept the estimates of the Bondi accretion rate 
and jet power at face value in what follows. Clearly it will be important
to see how these observational constraints improve with future X-ray 
missions. In section \ref{sec:rapidexpansion}, we consider the impact
of uncertainties in the calculation of the jet power from the physical properties of cavities observed in the Chandra data.

\begin{figure}
\centering
\includegraphics[scale=0.7]{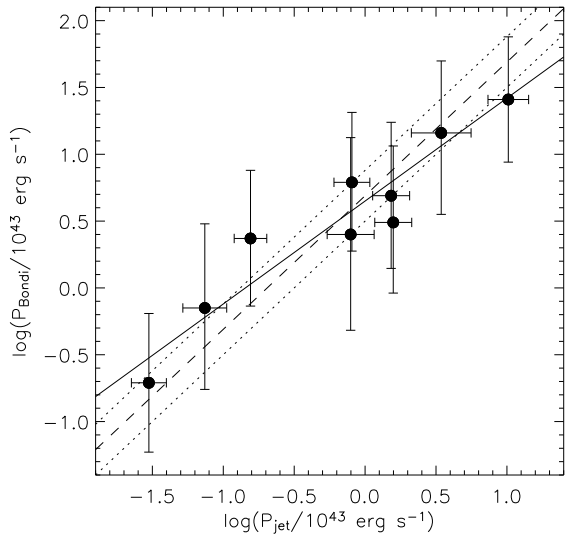}
\caption{The empirical relationship between the Bondi accretion power ($P_{\rm Bondi}=0.1 \dot{M}_{\rm Bondi} c^2$) and jet power ($P_{\rm jet}$) for nine nearby, X-ray luminous giant elliptical galaxies derived by A06. The error bars 
plotted include a systematic uncertainty of 0.46 dex in $\log P_{\rm Bondi}$.
The fitted power-law model predicted by our jet models ($A=0.69 \pm 0.19$ and $B=1$) is represented by the dashed (best-fit) and dotted (error bars) lines; the best-fit power-law model determined by A06 is shown as the solid line. }
\label{fig:comp}
\end{figure}

The A06 results for the nine systems are shown in Figure
\ref{fig:comp},  where $P_{\rm Bondi}$ is the total accretion
power released for an efficiency of 10\%, $P_{\rm Bondi}=0.1
\dot{M}_{\rm Bondi} c^2$. The error bars plotted include a systematic
uncertainty of 0.46 dex in $\log P_{\rm Bondi}$, implied by the
intrinsic dispersion of 0.23 dex in the $\log M_\bullet - \log \sigma$
relation (see A06).

The correlation between $\dot{M}_{\rm Bondi}$ (or equivalently, $P_{\rm Bondi}$) and jet power $P_{\rm jet}$ was expressed by A06 as a power-law model of the form
\begin{equation} \label{eq:powerlaw}
\log \frac{P_{\rm Bondi}}{10^{43} \, {\rm erg \; s}^{-1}} = A + B \log \frac{P_{\rm jet}}{10^{43} \, {\rm erg \; s}^{-1}},
\end{equation}
with $A=0.65 \pm 0.16$ and $B=0.77 \pm 0.20$. This power-law model is shown in Figure \ref{fig:comp} as a solid line. 

\section{$\dot{M}$--$P_{\rm jet}$ relation: implications for black hole spin and accretion rate} \label{sec:constr}

The A06 results are interesting in two regards: First, they show
that the central AGNs in these systems are extremely sub-Eddington;
that is, they have extremely low accretion rates compared to the
Eddington rate, $\dot{M}_{\rm Bondi}/\dot{M}_{\rm Edd} \lesssim
10^{-3}$, where $\dot{M}_{\rm Edd} \equiv 22 M_\bullet /(10^{9}
M_\odot) \; M_\odot \, {\rm yr}^{-1}$.  Our ADAF model for the
accretion flow is only valid for such highly sub-Eddington flows.
Second, as noted by A06, the correlation implies that a non-negligible
fraction ($P_{\rm jet}/(\dot{M}_{\rm Bondi} c^2) = 2.2^{+1.0}_{-0.7}
\%$ for $P_{\rm jet} = 10^{43} \, {\rm erg \; s}^{-1}$) of the energy
associated with the rest mass of the gas entering $r_A$ is channeled
into jet power.

\subsection{Reconsidering the mass accretion rate} \label{sec:massaccrete}

Presumably less matter than the amount predicted by the Bondi rate gets down to the black hole for several reasons. As the gas in an ADAF has angular momentum, accretion is driven not just by gravity as in the Bondi flow but also by the rate of angular momentum transport characterized by the parameter $\alpha$ in our ADAF model;  given a certain ambient density in the external medium, the accretion rate predicted by the ADAF model is lower than the Bondi accretion rate: $\dot{M}_{\rm ADAF} \sim \alpha \dot{M}_{\rm Bondi}$ (e.g., N98; \citealt{proga03}), with $\alpha < 1$. Furthermore, some part of the gas may be prevented from being accreted due to winds and/or convection (e.g., BB99; \citealt{quataert00,proga03,igu03}) occuring in the ADAF, reducing even more $\dot{M}_{\rm ms}$ compared to $\dot{M}_{\rm Bondi}$.  We allow $\dot{M}_{\rm ms}$ to be smaller than $\dot{M}_{\rm Bondi}$ by introducing the parameter $\epsilon_{\rm Bondi}$  such that  $\dot{M}_{\rm ms} = \epsilon_{\rm Bondi} \dot{M}_{\rm Bondi}$. The parameter $\epsilon_{\rm Bondi}$ represents the fraction of material supplied by the external medium at the Bondi radius that ultimately reaches the innermost stable circular orbit of the accretion flow and gets accreted afterwards. Therefore this parameter encompasses our ignorance about the possible physical processes that may modify the density profile of the accretion flow with respect to the ADAF solution.  We note that here we are only considering the simplest situation where the density profile is affected.  Winds too will reduce the rate of energy and angular momentum accretion onto the central black hole. We defer the discussion of such complications to \textsection \ref{sec:winds}.

Including this modification into our models gives jet power relations of the form 
\begin{equation} \label{eq:power}
P_{\rm jet} (P_{\rm Bondi}, \alpha, \epsilon_{\rm Bondi}, j) \propto P_{\rm Bondi}.
\end{equation}
This relationship has the same functional form as the correlation found by A06 (Equation \ref{eq:powerlaw}) and we can now use the above equation to constrain the main model parameters ($\alpha$, $\epsilon_{\rm Bondi}$ and $j$) using A06 results.

Comparing Equations \ref{eq:powerlaw} and \ref{eq:power}, it follows
that our models predicts the slope $B=1$, which is somewhat higher
than the value obtained by A06, $B=0.77 \pm 0.20$, although the
difference is not statistically significant (see
Fig. \ref{fig:comp}). As $A$ and $B$ in Equation \ref{eq:powerlaw}
presumably are strongly correlated, we fit A06's data to a
power-law model with the fixed value $B=1$ to find the corresponding
value of $A$. Using the $\chi^2$ fit statistics, which accounts only
for errors in the values of $P_{\rm Bondi}$, we find $A=0.69 \pm 0.19$
with $\chi^2=2.4$ for 8 degrees of freedom, which indicates that our
$B=1$ fit also provides a good description of the data, and hints that 
the systematic errors in the data points are not fully independent. 
The power-law
model with $A=0.69$ and $B=1$ is plotted in Figure \ref{fig:comp} as
the dashed line; the dotted lines delimitate the corresponding statistical
uncertainty in this power-law.

In the $B=1$ model, the value of the parameter $A$ is a measure of the
jet efficiency $\eta_{\rm jet}$.  Setting $B=1$ in
Eq. \ref{eq:powerlaw} we have
\begin{equation} \label{eq:A}
A = \log \left( \frac{P_{\rm Bondi}}{P_{\rm jet}} \right) 
= \log \left( \frac{0.1}{\eta_{\rm jet} \epsilon_{\rm Bondi}} \right),
\end{equation} 
which can be recast as 
\begin{equation} \label{eq:eta}
\log(\eta_{\rm jet})=-(1+A+\log \epsilon_{\rm Bondi}).
\end{equation} 

\begin{figure}
\centering
\includegraphics[scale=0.75]{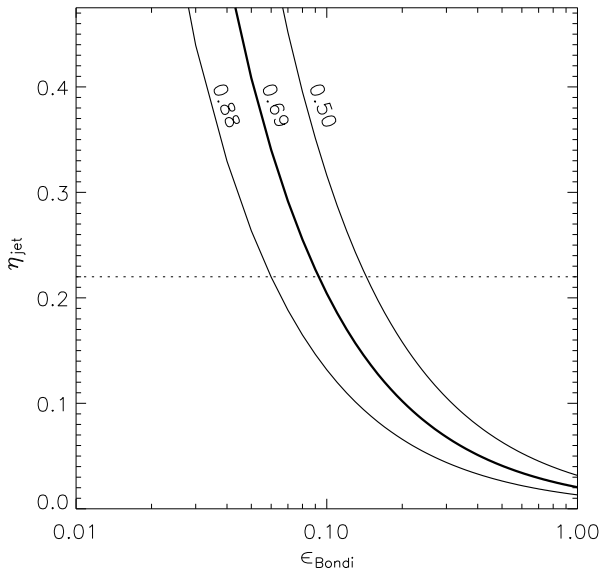}
\caption{Relation between the parameters $\eta_{\rm jet}$ and $\epsilon_{\rm Bondi}$ required to reproduce the observed correlation between $\dot{M}_{\rm Bondi}$ and $P_{\rm jet}$. Each line correspond to a different value of $A$ (shown above the lines), the thicker line represents the best-fit value of $A$.  We limit the value of efficiency to the theoretical maximum for the hybrid model with $\alpha = 0.04$ ($48 \%$).   The dotted line corresponds to the maximum efficiency for the Blandford-Znajek model with $\alpha = 0.04$.  These correspond to maximally spinning black holes with $j=0.998$. The inferred efficiencies decline with increasing $\alpha$ since the plausible values of $\epsilon_{\rm Bondi}$ increase.  For $\alpha=0.3$, the maximal efficiencies are $\eta=0.07$  (BZ model) and $0.24$ (hybrid model).}
\label{fig:eta}
\end{figure}

Figure \ref{fig:eta} shows the jet efficiency factor corresponding to
the observed value of $A=0.69 \pm 0.19$ ($B=1$), as a function of the mass
accretion parameter $\epsilon_{\rm Bondi}$.  The value of $\eta_{\rm
jet} \approx 2\%$ derived by A06 is, in fact, the minimum value and
corresponds to $\epsilon_{\rm Bondi}=1$ or $\dot{M}_{\rm ms} =
\dot{M}_{\rm Bondi}$. Since $\epsilon_{\rm Bondi}$ is expected to be 
significantly less than unity, the required jet efficiency will be higher.

Since in our models for the jet power, the jet efficiency factor is a
function of the viscosity parameter $\alpha$ and the black hole spin
$j$, the above equations imply that the value of the parameter $A$ is
a function of the parameters $\alpha$, $\epsilon_{\rm Bondi}$ and $j$.
And therefore, for a given value of $\alpha$, we can use the observed
value of $A=0.69 \pm 0.19$ to explore the range of possible values of
$\epsilon_{\rm Bondi}$ and $j$. 
As described previously, various
different analysis and arguments suggest that $\alpha$ ranges between
0.04 and 0.3.  In the discussion below, we will derive results for the
two values that bound the range: $\alpha=0.04$ and $0.3$.

\begin{figure*}
\centering
\begin{minipage}[b]{0.48\linewidth}
\includegraphics[width=\linewidth]{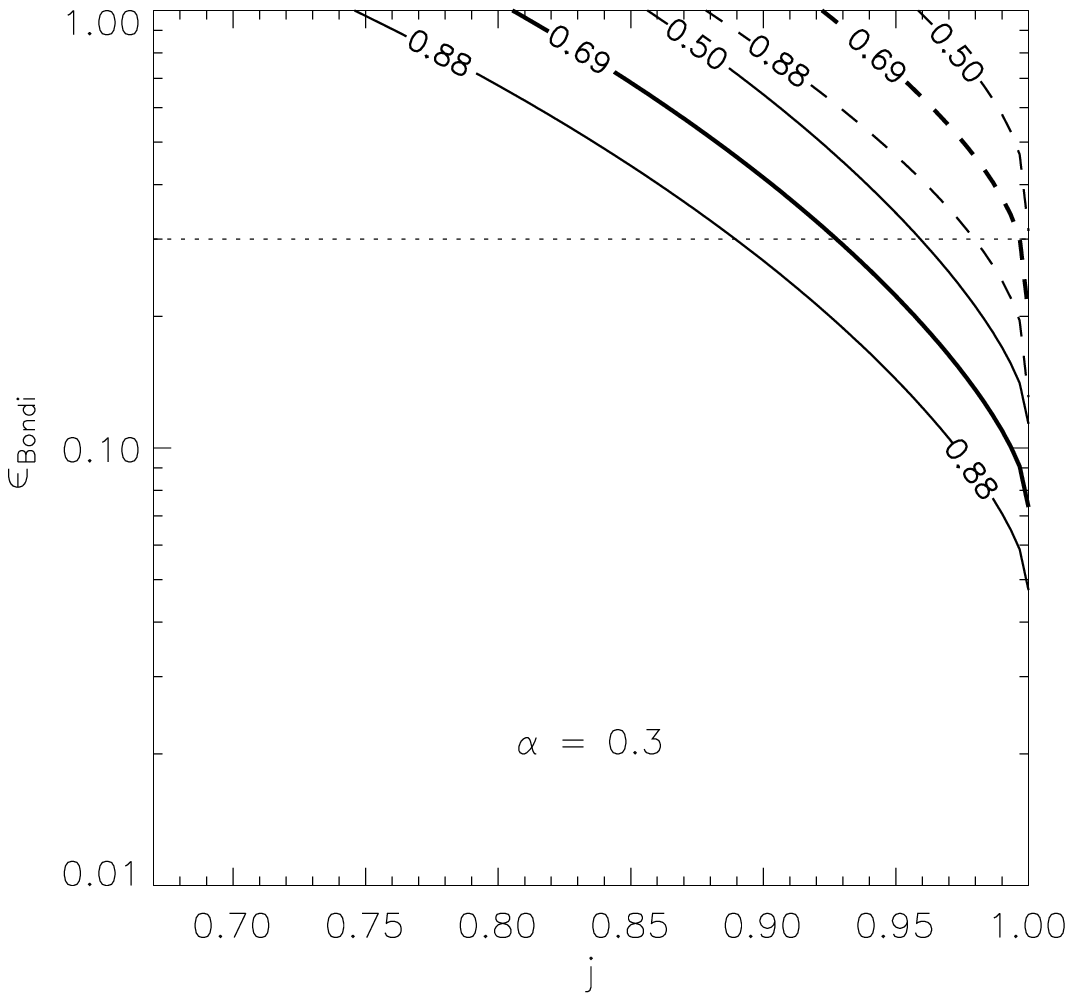}
\end{minipage}
\hfill
\begin{minipage}[b]{0.48\linewidth}
\includegraphics[width=\linewidth]{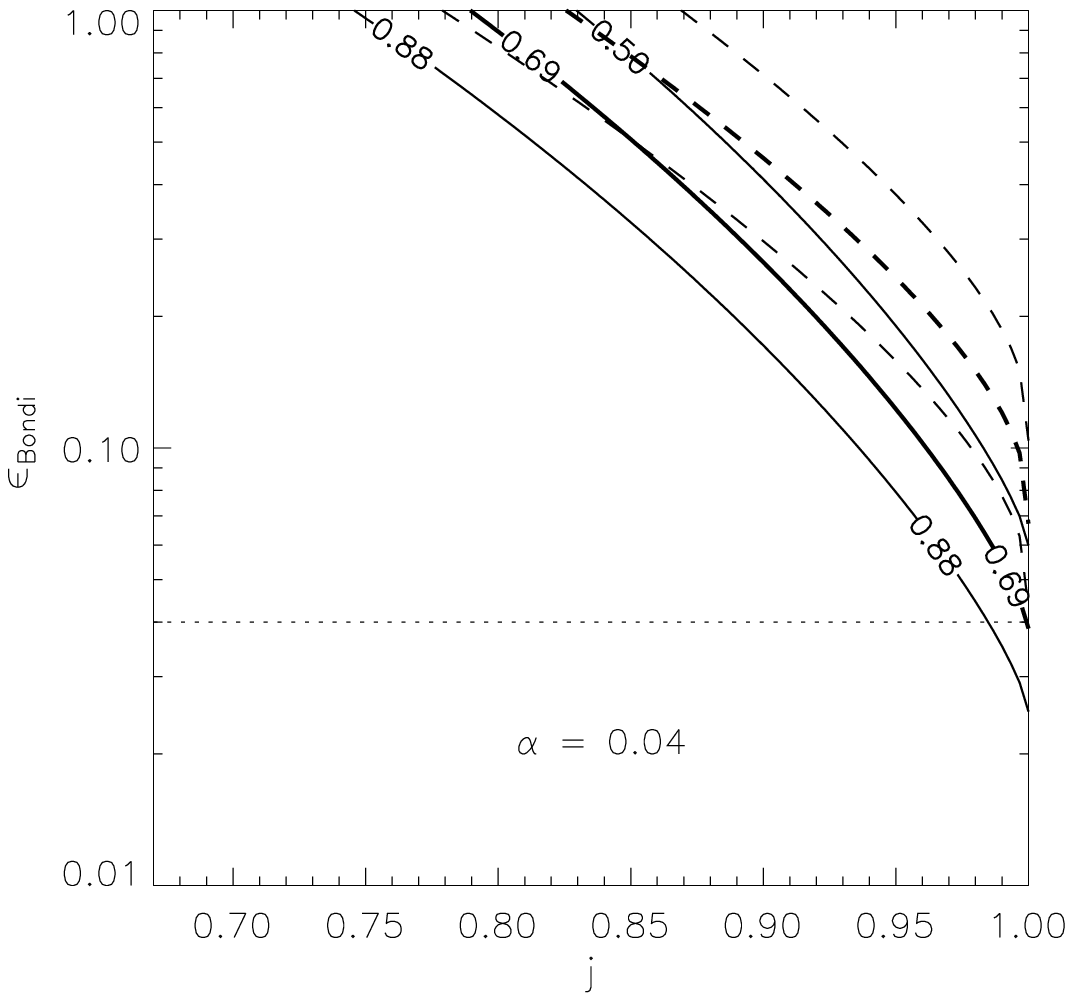}
\end{minipage}
\caption{Range of values for the parameters $\epsilon_{\rm Bondi}$ and $j$ from the jet models, which reproduce the measured values of $A$ of the empirical correlation between $\dot{M}_{\rm Bondi}$ and $P_{\rm jet}$ (Equation \ref{eq:A}).  The dashed lines show the predictions of the Blandford-Znajek model and the solid lines show those of the hybrid model. The left panel corresponds to $\alpha=0.3$ and the right panel to $\alpha=0.04$, the two values that bound the range of plausible values for the viscosity parameter. The label beside each line is the value of $A$ for that line. The thickest lines correspond to the best fitting value of $A$. The horizontal dotted lines indicate the value $\epsilon_{\rm Bondi}=\alpha$ (i.e., $\dot{M}_{\rm ms} = \dot{M}_{\rm ADAF}$).}
\label{fig:contour}
\end{figure*}

Figure \ref{fig:contour} shows the parameter space ($j$,
$\epsilon_{\rm Bondi}$) corresponding to the observed values of $A$
for the BZ (dashed line, \textsection \ref{sec:bz}) and the hybrid
(solid line, \textsection \ref{sec:disk}) jet models. The two panels
show the results for the two different values of $\alpha$ near the
innermost stable circular orbit, $\alpha=0.3$ ($\beta \sim 1$, left
panel) and $\alpha=0.04$ ($\beta \sim 10$, right panel).  We remind
the reader that $\beta$ is the ratio of the gas pressure to the
magnetic pressure near the radius of the marginally stable orbit.  The
label beside each contour is the value of $A$ for that line. We show the
effect of considering an uncertainty of $\pm 0.19$ in $A$. 
Note that the range we have adopted assumes that the uncertainty in each 
data point is independent. $A$ would be smaller if, for example,
the jet power was systematically under-estimated in all systems 
(see \S\ref{sec:rapidexpansion}).

Based on the arrangement of the contours in this figure, we find that the observed tight correlation between accretion rates and jet powers implies a narrow range of black hole spins for the elliptical galaxies of the sample of A06, irrespective of the value of $\alpha$ and the specific model adopted, with the main result being that the central black holes powering the jets in these systems must be rapidly spinning.   Specifically, we find that if $\alpha \approx 0.3$ (Figure \ref{fig:contour}a), as required to ensure an agreement between ADAF model predictions for XRBs and the observations \citep{quataert99}, then the hybrid model requires $j \gtrsim 0.75$, while the BZ model implies $j \gtrsim 0.88$.   If the value of the viscosity parameter is on the low side, $\alpha \approx 0.04$ (Figure \ref{fig:contour}b), both jet models require $j \gtrsim 0.75$.   In all cases, this corresponds to $\eta_{\rm jet} \gtrsim 2 \%$.  

These lower limits for both $j$ and $\eta_{\rm jet}$ correspond to the situation where rate of mass accretion onto the black hole equals the Bondi rate ($\epsilon_{\rm Bondi} \sim 1$).   As discussed above, only a fraction of the Bondi rate is likely to make its way to the black hole and realistically, $\dot{M}_{\rm ms} \lesssim \dot{M}_{\rm ADAF}$.   This restricts the allowed region of parameter space in Figure \ref{fig:contour} to that below the horizontal dotted lines ($\epsilon_{\rm Bondi}=\alpha$) corresponding to $\dot{M}_{\rm ms} = \dot{M}_{\rm ADAF}$.     

Subject to these constraints on the mass accretion rate, the observed
correlations between mass accretion rate and jet power imply that in
the case of the hybrid model, the central black holes must have spins
$j \gtrsim 0.89$ if $\alpha \approx 0.3$.  The corresponding range for
the mass accretion rate onto the black hole is $\dot{M}_{\rm ms}
\approx (0.05-0.3) \dot{M}_{\rm Bondi} \approx (0.16-1.0) \dot{M}_{\rm
ADAF}$.  If $\alpha$ is on the low side, then the black holes in the
hybrid model must be spinning rapidly ($j \gtrsim 0.98$) and the
corresponding range for the mass accretion rate is $\dot{M}_{\rm ms}
\approx (0.02-0.04) \dot{M}_{\rm Bondi} \approx (0.5-1.0) \dot{M}_{\rm
ADAF}$.  For a given value of $j$, the
corresponding jet efficiency factor can be read off from Figure
\ref{fig:eta}.  For example, if $\alpha = 0.3$ and $j=0.89$, 
$\eta_{\rm jet} = 4.5\%$, giving an overall efficiency of 
$\eta_{\rm jet}\epsilon_{\rm Bondi} = 1.3\%$.

In the case of the BZ model, the spin distribution corresponds to $j \gtrsim 0.98$ ($\alpha=0.3$) and $j > 1$ ($\alpha=0.04$).   The latter is unphysical and indicates that values of viscosity parameters as low as $\alpha=0.04$ are disfavored.  For this line of reasoning, we find that the BZ model requires $\alpha \geq 0.2$ in order to guarantee $j \leq 1$ and $\dot{M}_{\rm ms} \leq \dot{M}_{\rm ADAF}$.   It is also possible these ``problems'' reflect the limitations of the BZ model.  As indicated previously, the BZ model is, strictly speaking, applicable only in the limit of slow rotation where the unipolar approximation is reasonable.   For this reason, and for the clarity of the discussion, we shall hereafter only discuss the results in the context of the hybrid model though we shall continue to show the results for both models.

Since our results indicate high spin values for jet-emitting black
holes, it useful to bear in mind that recent numerical relativistic
MHD simulations of thick-disks of \citet{gammie04} suggest that the
accreting plasma will bring the black hole in spin equilibrium not at
the maximal value of $j=0.998$ \citep{thorne74} but at a lower value
of $j\approx 0.93$.  This is not a trivial difference.  At high spin
rates, a $7\%$ reduction in $j$ translates into a factor of $\sim 3$ reduction 
in the jet power.  In the specific case of the A06 elliptical galaxies, we
find that black holes in these galaxies must be fed at rates
$\epsilon_{\rm Bondi} \gtrsim 0.05$, depending on the value of
$\alpha$ and the mechanism of jet powering at work.  In the particular
case of a hybrid model with $\alpha \approx 0.3$, the black holes
cannot be fed at rates much lower than $\dot{M}_{\rm ADAF}$ (i.e,
$0.2 \lesssim \epsilon_{\rm Bondi} \lesssim 0.3$, where the lower
limit is due to the reduced efficiency associated with $j=0.93$ and
corresponds to $\dot{M}_{\rm ms} = \dot{M}_{\rm ADAF}$), limiting
the potential role of mass loss through winds.

\subsection{Reconsidering the jet power estimates} \label{sec:rapidexpansion}

A06 uses the X-ray observations of the intracluster medium cavities to estimate the jet power.
To derive the energy $E$ required to create the observed bubbles in the X-ray emitting gas, A06 assume that the cavities are inflated slowly and obtain the relationship $E=4PV$ for $\gamma = 4/3$ (relativistic plasma), where $P$ is the thermal pressure of the surrounding X-ray emitting gas, $V$ is the volume of the cavity and $\gamma$ is the mean adiabatic index of the fluid within the cavity.   This is the minimum energy required to inflate the cavities and does not take into account any additional energy that may have gone into heating the intracluster medium.    A more realistic scenario is likely to involve overpressurized bubbles, which upon injection expand rapidly to reach pressure equilibrium with their surroundings, and in the process generate weak shock waves that heat the X-ray emitting gas.   \citet{nusser06} calculate the total energy $E'$ deposited by the jets in such cases and find that
\begin{equation} \label{eq:nusser}
E' \approx 3 PV \left( \frac{P_i}{P} \right)^{1/4}, 
\end{equation} 
where $P_i$ is the thermal pressure of the surrounding gas at the location where the bubble is injected.  Equation \ref{eq:nusser} suggests that if a bubble is injected with overpressure $P_i/P \gtrsim 10$ then $E' \gtrsim 6 PV$. We consider the possibility that as an extreme case the bubble energies are twice the value assumed by A06 ($E'=8PV$), and calculate the impact of this on the observed jet powers and on the results derived from our models.

Following A06, we calculate the modified jet powers as $P'_{\rm jet} = E'/t_{\rm age} = 2 P_{\rm jet}$ and refit the power-law model with $B=1$ to the modified data, taking into account the proper error propagation on the values of $P'_{\rm jet}$. We obtain $A = 0.39 \pm 0.19$ with $\chi^2=2.4$ for 8 degrees of freedom. As the assumed value of $E'$ implies higher jet powers, our models need higher values of $j$ to reproduce the increased values of $P'_{\rm jet}$. In particular, if $\alpha \approx 0.3$ these values of $A$ imply $j_{\rm min} \approx 0.84$ for the hybrid model and $j_{\rm min} \approx 0.95$ for the BZ model. Therefore, a narrower range of still large spins is required to explain $P'_{\rm jet}$ and the implied numerical values of $\epsilon_{\rm Bondi}$ and $\epsilon_{\rm ADAF}$ also increase slightly.

One important uncertainty in the above calculation (and also in that of A06) is the estimate of the age of the bubbles. A06 calculate the ages using the formula $t_{\rm c_s}=D/c_s$, where $D$ is the distance of the bubble centre from the black hole and $c_s$ is the adiabatic sound speed, but as discussed by \citet{birzan04} (see also \citealt{rafferty06}) there are two other ways of estimating the age of the cavities (Equations 3 and 4 of \citealt{birzan04}) which result in longer timescales when compared to $t_{\rm c_s}$.   An underestimate of the timescales involved impacts the estimate of jet power, as determined above, by artificially enhancing the jet power above the \qm{true} value.   To take this into account, we consider the possibility that the A06 ages are too low by a factor of $\sim 2$ \citep{birzan04,rafferty06}, such that the modified age is $t_{\rm age} \approx 2 t_{\rm c_s}$, implying a smaller jet power $P''_{\rm jet} = 1/2 P_{\rm jet}$.   With reference to the original jet power estimates of A06, we find that revising the timescales as suggested here results, not surprisingly, in a systematic decrease in lower limits of $j$-values.   For example, in the case of the $\alpha=0.3$ hybrid model, we find that $j_{\rm min} \approx 0.6$.

In spite of the above-mentioned concerns with the A06 estimates of the total jet energy and the associated timescales,  it seems likely that their end-product, the estimates of the jet power in the nine elliptical galaxies, is  reasonable. The impact of underestimating the total jet energy very nearly cancels out the impact of underestimating the timescales involved and the analysis that we have presented thus far will only be minimally affected.  

\subsection{Reconsidering ADAF models: the effect of accretion disk winds} \label{sec:winds}

Finally, we have alluded previously to the possibility that the
accretion flow may be modified by mass loss in the form of winds.
Winds from the accretion flow are
likely to be the norm rather than the exception (e.g., \citealt{blandford05}) and these winds
will remove mass, angular momentum and energy from the flow.  As a
consequence, ADAF-like accretion flows with winds (advection-dominated inflow-outflow solution, ADIOS model, BB99)
will have a different dynamical structure than ADAFs without winds.
For instance given the same value of $\dot{M}$ near the black hole,
the angular velocity, scale height, total pressure and magnetic field
strength predicted by the ADAFs with and without winds will be
different.  And since our models for the jet power are linked to the
accretion flow model through the quantities $H$, $\Omega'$ and $B$, it
is not inconceivable that jet powers linked to no-wind ADAFs could be very
different from jet powers linked to ADIOS models.

The ADIOS solution has three parameters in addition to those characterizing the no mass loss ADAF model: $p_w$, $\lambda_w$ and $\epsilon_w$. These parameters describe how much mass, energy and angular momentum, respectively, the wind removes from the accretion flow and specific choice of values for these three parameters leads to different types of winds (for more information see BB99).  To assess the implications for jet power if the true underlying model is more correctly described by the ADIOS solution, we begin by fixing the parameters common to the two models (the ADIOS and the no-wind model) as follows:  $\alpha = 0.1$, $\gamma=1.5$, and $M_\bullet=10^9 M_\odot$ (we have confirmed that varying these particular variables does not affect our conclusions).   We then consider various ADIOS solutions arising from varying the wind parameters across the following range: $p_w=0-1$, $\lambda_w=0.1-0.75$ and $\epsilon_w=0.1-0.5$. This range of values encompasses all the interesting types of winds.  

We found that while some features of the two models are very different, the solutions for $H$, $\Omega'$ and $B$ at the radius of marginally stable orbit in the ADIOS model are, to first order, comparable to the corresponding solutions derived from the no-wind model.  In detail, the differences are such that the ADIOS jet powers are always smaller than the no-wind values by factor of order unity, regardless of the combination of values of the wind parameters.  Lower jet powers in the ADIOS model, in turn, imply both a narrower range of spins for jet-emitting black holes and that these black holes must be spinning faster than those embedded in the corresponding no mass loss ADAF in order to reproduce the correlations derived by A06. In other words, if winds are indeed the norm, then the central black holes must be spinning even more rapidly than suggested by our analysis above.   

\subsection{Comparison with previous estimates of jet efficiency}

A06 reported an efficiency of conversion of $\dot{M}_{\rm
Bondi}$ into jet power of $P_{\rm jet}/\dot{M}_{\rm Bondi} c^2 \approx
2 \%$ for $P_{\rm jet} = 10^{43} \, {\rm erg \; s}^{-1}$. In our
modelling, this corresponds to the case where $\epsilon_{\rm Bondi}=1$
(i.e. $\dot{M}_{\rm ms} = \dot{M}_{\rm Bondi}$).  Since it is more
likely that $\dot{M}_{\rm ms} \lesssim \dot{M}_{\rm Bondi}$, our
models suggest that the efficiency of conversion of the accreting
matter into jet power is considerably higher.  For instance,
$\eta_{\rm jet}$ may reach values as high as $\approx 50\%$ for high
spin rates.  The extraction of spin energy from the holes is
responsible for this noticeable increase in the jet efficiency. We
note that the increase in the jet power with black hole spin has been
verified in numerical simulations of jets (e.g.,
\citealt{devilliers05, mckinney05, hawley06}).

The uppper limit we have obtained for $\eta_{\rm jet}$ is more than an order of magnitude larger than that  reported by \citet{armitage99}. This result is mainly due to the fact that we have included a Kerr metric shear-driven dynamo \citep{meier99}, which enhances the field strength and was not included by these authors (see \textsection \ref{sec:riaf} and Appendix).   Based on the calculations of \citet{armitage99},  \citet{cao04} have recently asserted that accretion flow-jet models of the kind we have considered cannot account for the jet powers of  a sample of low accretion rate 3CR FR I radio galaxies observed using the Hubble Space Telescope (jet powers in the range $\sim 10^{41}-10^{45} \, {\rm erg \; s}^{-1}$).   As a result of much higher efficiencies, adopting the critical accretion rate $\dot{M}_{\rm crit} \sim \alpha^2 \dot{M}_{\rm Edd}$  \citep{esin97} above which the ADAF solution ceases to be valid, and taking $j=0.998$ and $\alpha=0.3$, our jet models yield $P_{\rm jet} \approx 10^{45-46} \, {\rm erg \; s}^{-1}$, which is more than enough to account for the observed jet power in the sample used by  \citet{cao04}.

The adopted model for the jet-formation mechanism predicts a linear relation between accretion rate and jet power, while the measured best-fit slope of the relation measured by A06 is marginably better fit by a non-linear relationship.  More than likely, this marginal discrepancy is due to the small number of systems in the A06 sample.   However, we note that if $\epsilon_{\rm Bondi}$ has a slight dependence on $j$, as suggested by recent numerical simulations of \citet{hawley06}, then $\epsilon_{\rm Bondi}$ will also exhibit a weak dependence on $P_{\rm jet}$ and the agreement of the model with the observed correlation will be  improved. We verified that if $\epsilon_{\rm Bondi} \propto P_{\rm jet}^{0.2}$ the slope predicted by the model agrees with the observed best-fit slope.  In the numerical simulations, the dependence of $\dot{M}_{\rm ms}$ on $j$ arises due amplified magnetic fields that act to transfer angular momentum from the hole to the accretion disk.   In addition, it is not inconceivable that the dependence $\epsilon_{\rm Bondi}(P_{\rm jet})$ might also emerge as a consequence of feedback effects of the jet on the material fueling the accretion flow.

\section{Conclusions} \label{sec:conclusion}

We have employed two physical models for the black hole-accretion
flow-jets with features that are broadly consistent with the results
of numerical simulation to understand the empirical
correlation between accretion rates and jet powers of X-ray luminous
elliptical galaxies derived by A06: the classical Blandford-Znajek
model and a hybrid model proposed by \citet{meier01}. In the BZ model the energy is extracted electromagnetically from the spinning black hole; in the hybrid model the magnetic field threads both the inner and the outer regions
of the accretion flow and the energy that powers the outflow is
extracted from both the rotation of the disk as well as from the
rotation of the black hole, via the frame-dragged accretion flow inside the
hole's ergosphere.  We assume that the accretion flow is
advection-dominated (ADAF) and take into account general relativistic
effects not fully appreciated before in these models. 

In the absense of disk winds and feedback, the model suggests that the 
jet power should be linearly dependent on the accretion mass flow
rate. The normalisation of this relation is dependent on the disk
viscosity and the black hole spin, with the dispersion around the best-fit 
correlation of A06 being caused by different values of the black hole 
spin, $j$, and the ratio of the disk mass flow rate to the
accretion rate onto the black hole, $\epsilon_{\rm Bondi}$. We find that the jet efficiency ($\eta_{\rm jet}=P_{\rm jet}/\dot{M}_{\rm ms} c^2$) can exceed $10 \%$ so that the jet may carry away an appreciable fraction of the rest mass energy of the accreted matterial.

We compared our jet power models to the jet power estimates made by A06. Adopting 
typical values of the viscosity parameter
$\alpha \sim 0.04 - 0.3$, the $\dot{M}_{\rm Bondi}$ vs. $P_{\rm
jet}$ correlation implies a narrow range of spins $j \approx 0.75-1$
and accretion rates $\dot{M}_{\rm ms} \approx (0.04 - 1) \dot{M}_{\rm
Bondi}$. 
If we further demand that the mass accretion rate be restricted to
$\dot{M}_{\rm ms} \lesssim \dot{M}_{\rm ADAF} \sim \alpha \dot{M}_{\rm
Bondi}$, as is both likely to be the case and consistent with ADAF
model expectations, we find that the observed correlations require $j
\gtrsim 0.9$; i.e., the correlations imply rapidly spinning black
holes in all of the target galaxies.  If additionally, the ADAF 
accretion flows also experience
mass, energy and angular momentum loss via winds as in the ADIOS model
proposed by BB99, the correlations indicate nearly-maximally spinning
black holes (see \textsection \ref{sec:winds}).  
It is reassuring that semi-analytic cosmological simulations of the spin evolution of black holes through mergers and gas accretion \citep{volonteri05} and estimates of the radiative efficiencies of global populations of quasars based on Soltan-type arguments (e.g., \citealt{soltan82,yu02,wang06}) suggest that most nearby massive holes are rapidly rotating. 

The relatively small scatter in the correlation between $P_{\rm
Bondi}$ and $P_{\rm jet}$ of A06, combined with the strong dependence
of jet power on black hole spin implies large black hole spins is
probably a general result valid for all elliptical galaxies. Of course the
sample considered by A06 is relatively small, 
but if this result holds-up in larger samples, and can be shown to also apply 
to the larger central radio galaxies of galaxy clusters 
\citep{rafferty06}, 
our results suggest that they have the most powerful jets 
because they have sufficient black hole mass to host relatively high
mass flow rates while the disk remains in the ADAF state (and hence
is able to generate strong poloidal magnetic fields near the black hole). 
In this picture, all central ellipticals have the capacity to produce
powerful jets, and the resulting jet power is determined by the structure
of the accretion disk.
Our results reveal a potentially fundamental connection between
black holes and the formation of the most massive galaxies: the
central holes in the cores of galaxy clusters must be rapidly
rotating, in order to make the radio jets powerful enough to provide
an effective feedback mechanism and quench the cooling flows,
therefore preventing star formation and explaining the observed galaxy
luminosity function (e.g., \citealt{croton06,bower06}) as well as
accounting for the observed X-ray luminosity -- temperature and X-ray luminosity -- Sunyaev-Zeldovich effect correlations for the intracluster medium (e.g., \citealt{babul02,mccarthy04}).

\section*{Acknowledgments}

We are very grateful to David L. Meier and Chris Done for their 
valuable guidance and discussions. RSN thanks Teddy Cheung, Feng Yuan and Xinwu Cao for
helpful discussions and acknowledges the hospitality of the Institute
for Computational Cosmology (ICC, Durham University) where part of this
work was carried out. RSN and TSB acknowledges support from the
Brazilian institutions CNPq, CAPES and FAPERGS. RGB thanks PPARC for
the support of a senior research fellowship. AB thanks the ICC for hospitality
shown to him during his tenure there as Leverhulme Visiting Professor,
and is deeply appreciative of support from the Leverhulme Trust as
well as NSERC (Canada). This work was supported by the European
Commission’s ALFA-II programme through its funding of the
Latin-american European Network for Astrophysics and Cosmology
(LENAC).

\appendix

\section{Derivation of the jet power}

We list the equations we used to compute the dependence of the jet power on $\alpha$, $j$ and $\dot{M}$ using the Blandford-Znajek model (\textsection \ref{sec:bz}) and the hybrid model (\textsection \ref{sec:disk}). The jet power is given by Equations \ref{eq:powerbz} (BZ model) and \ref{eq:powerdisk} (hybrid model).  The code that implements the equations described in this work and returns the jet power is available at the URL http://www.if.ufrgs.br/$\sim$rns/jetpower.htm.

The following equations describe the self-similar ADAF structure \citep{narayan95}, where we use the black hole mass in solar units ($m=M_\bullet/M_\odot$), accretion rates in Eddington units ($\dot{m}=\dot{M}/\dot{M}_{\rm Edd}$, $\dot{M}_{\rm Edd}$ is the Eddington accretion rate defined in \textsection \ref{sec:correl}) and radii in Schwarzschild units ($r=R/(2 GM_\bullet/c^2 )$):
\begin{gather}
\Omega' = 7.19 \times 10^4 c_2 m^{-1} r^{-3/2} \; {\rm s}^{-1}, \\
B = 6.55 \times 10^8 \alpha^{-1/2} (1-\beta)^{1/2} c_1^{-1/2} c_3^{1/4} m^{-1/2} \dot{m}^{1/2} r^{-5/4} \; {\rm G}, \\
H/R \approx (2.5 c_3)^{1/2}.
\end{gather} 
The constants $c_1$, $c_2$ and $c_3$ are defined as
\begin{gather*}
c_1 = \frac{5+2\epsilon'}{3 \alpha^2} g'(\alpha,\epsilon'), \\
c_2 = \left[  \frac{2 \epsilon' (5+2\epsilon')}{9 \alpha^2} g'(\alpha,\epsilon') \right] ^{1/2}, \\
c_3 = c_2^2/\epsilon', \\
\epsilon' \equiv \frac{1}{f} \left( \frac{5/3-\gamma}{\gamma-1} \right), \\
g'(\alpha,\epsilon') \equiv \left[ 1+ \frac{18 \alpha^2}{(5+2\epsilon')^2} \right]^{1/2}. 
\end{gather*} 

The relations among $\alpha$, $\gamma$ and $\beta$ are given in \textsection \ref{sec:riaf}. 
The angular velocity of the field seen by an outside observer at infinity in the Boyer-Lindquist frame is $\Omega = \Omega' + \omega$, where the angular velocity of the local metric is given by \citep{bardeen72}
\begin{equation}
\omega \equiv -\frac{g_{\phi t}}{g_{\phi \phi}} = \frac{2 a M_\bullet}{a^2(R+2M_\bullet) + R^3},
\end{equation} 
using geometrized units ($G=c=1$). 

We estimate the field-enhancing shear caused by the Kerr metric following \citet{meier01} as $g=\Omega/\Omega'$, such that the azimuthal component of the field is given by $B_\phi = g B$ (see \textsection \ref{sec:riaf}). The poloidal component is related to the azimuthal component following \citet{livio99} as $B_p \approx H/R \; B_\phi \approx B_\phi$.

Lastly, we insert all the quantities defined above into Equations \ref{eq:powerbz} and \ref{eq:powerdisk}, and then evaluate the resulting equations at the marginally stable orbit of the accretion disk $R_{\rm ms}$, given by \citep{bardeen72}
\begin{gather}
R_{\rm ms} = M_\bullet \left\lbrace 3 + Z_2 - \left[ (3-Z_1)(3+Z_1+2Z_2) \right]^{1/2}  \right\rbrace, \\
Z_1 \equiv  1 + (1-j^2)^{1/3} \left[ (1+j)^{1/3} + (1-j)^{1/3} \right], \nonumber \\ 
Z_2 \equiv  (3 j^2 + Z_1^2)^{1/2}. \nonumber
\end{gather} 
Taking $R=R_{\rm ms}$ in the BZ model corresponds to assume that the strength of the magnetic field at the horizon of the hole is very similar to the corresponding strength at the marginally stable orbit. This is a reasonable assumption according to recent numerical simulations of jet formation (e.g., \citealt{hirose04,mckinney04}).

\bsp

\label{lastpage}


\begin{thebibliography}{}
\bibitem[\protect\citeauthoryear{Allen et al.}{2006}]{allen06} Allen, S. W., R. J. H Dunn, Fabian, A. C., Taylor, G. B.,  Reynolds, C. S. 2006, MNRAS, 372, 21 (A06)
\bibitem[\protect\citeauthoryear{Armitage \& Natarajan}{1999}]{armitage99} Armitage, P. J.,  Natarajan, P. 1999, ApJ, 523, L7
\bibitem[\protect\citeauthoryear{Babul et al.}{2002}]{babul02} Babul, A. 2002, MNRAS, 330, 329
\bibitem[\protect\citeauthoryear{Balbus \& Hawley}{1998}]{balbus98} Balbus, S. A.,  Hawley, J. F. 1998, Reviews of Modern Physics, 70, 1
\bibitem[\protect\citeauthoryear{Bardeen et al.}{1972}]{bardeen72} Bardeen, J. M., Press, W. H.,  Teukolsky, S. A. 1972, ApJ, 178, 347
\bibitem[\protect\citeauthoryear{Birzan et al.}{2004}]{birzan04} Birzan, L., Rafferty, D. A., McNamara, B. R., Wise, M. W.,  Nulsen, P. E. J. 2004, ApJ, 607, 800
\bibitem[\protect\citeauthoryear{Begelman et al.}{1984}]{begelman84} Begelman, M. C., Blandford, R. D.,  Rees, M. 1984, Rev. Mod. Phys., 56, 255
\bibitem[\protect\citeauthoryear{Blandford \& Znajek}{1977}]{bz77} Blandford, R. D.,  Znajek, R. 1977, MNRAS, 179, 433
\bibitem[\protect\citeauthoryear{Blandford}{2005}]{blandford05} Blandford, R. 2005, in Merloni A., Nayakshin S., Sunyaev R. A., eds, ESO Astrophysics Symposia, Growing Black Holes: Accretion in a Cosmological Context. Springer-Verlag, p. 477
\bibitem[\protect\citeauthoryear{Blandford \& Payne}{1982}]{bp82} Blandford, R. D.,  Payne, D. G. 1982, MNRAS, 199, 883
\bibitem[\protect\citeauthoryear{Blandford \& Begelman}{1999}]{bb99} Blandford, R. D.,  Begelman, M. 1999, MNRAS, 303, L1 (BB99)
\bibitem[\protect\citeauthoryear{Bondi}{1952}]{bondi52} Bondi, H. 1952, MNRAS, 112, 195
\bibitem[\protect\citeauthoryear{Bower et al.}{2006}]{bower06} Bower, R. G., Benson, A. J., Malbon, R., Helly, J. C.,
Frenk, C. S., Baugh, C. M., Cole, S., Lacey, C. G. 2006, MNRAS, 370, 645
\bibitem[\protect\citeauthoryear{Cao \& Rawlings}{2004}]{cao04} Cao, X.,  Rawlings, S. 2004, MNRAS, 349, 1419
\bibitem[\protect\citeauthoryear{Churazov et al.}{2002}]{churazov02} Churazov, E., Sunyaev, R., Forman, W.,  B\"ohringer, H. 2002, MNRAS, 332, 729
\bibitem[\protect\citeauthoryear{Churazov et al.}{2005}]{churazov05} Churazov, E., Sazonov, S., Sunyaev, R., Forman, W., Jones, C.,  B\"{o}hringer, H. 2005, MNRAS, 363, 91
\bibitem[\protect\citeauthoryear{Croton et al.}{2006}]{croton06} Croton, D. J., et al. 2006, MNRAS, 365, 11
\bibitem[\protect\citeauthoryear{Cui, Zhang \& Chen}{1998}]{cui98} Cui, W., Zhang, S. N.,  Chen, W. 1998, ApJ, 492, L53
\bibitem[\protect\citeauthoryear{Dalla Vecchia et al.}{2004}]{dv04} Dalla Vecchia, C., Bower, R. G., Theuns, T., Balogh, M. L., Mazzotta, P., Frenk, C. S. 2004, MNRAS, 355, 995
\bibitem[\protect\citeauthoryear{De Villiers et al.}{2005}]{devilliers05} De Villiers, J. P., Hawley, J. F., Krolik, J. H.,  Hirose, S. 2005, ApJ, 620, 878
\bibitem[\protect\citeauthoryear{Di Matteo et al.}{2001}]{dmt01b} Di Matteo, T., Johnstone, R. M., Allen, S. W.,  Fabian, A. C. 2001, ApJ, 550, L19
\bibitem[\protect\citeauthoryear{Di Matteo et al.}{2003}]{dmt03} Di Matteo, T., Allen, S. W., Fabian, A. C., Wilson, A. S.,  Young, A. J. 2003, ApJ, 582, 133
\bibitem[\protect\citeauthoryear{Esin et al.}{1997}]{esin97} Esin, A. A., McClintock, J. E.,  Narayan, R. 1997, ApJ, 489, 865
\bibitem[\protect\citeauthoryear{Fabian et al.}{2006}]{fabian06} Fabian, A. C., Sanders, J. S., Taylor, G. B., Allen, S.
W., Crawford, C. S., Johnstone, R. M., Iwasawa, K. 2006, MNRAS, 366, 417
\bibitem[\protect\citeauthoryear{Falcke et al.}{2004}]{falcke04} Falcke, H., K\"{o}rding, E.,  Markoff, S. 2004, A\&A, 414, 895
\bibitem[\protect\citeauthoryear{Ferrari}{1998}]{ferrari98} Ferrari, A. 1998, ARA\&A, 36, 539
\bibitem[\protect\citeauthoryear{Gallo et al.}{2003}]{gallo03} Gallo, E., Fender, R. P.,  Pooley, G. G. 2003, MNRAS, 344, 60
\bibitem[\protect\citeauthoryear{Gammie et al.}{2004}]{gammie04} Gammie, C. F., Shapiro, S. L.,  McKinney, J. C. 2004, ApJ, 602, 312
\bibitem[\protect\citeauthoryear{Ghosh \& Abramowicz}{1997}]{ghosh97} Ghosh, P.,  Abramowicz, M. A. 1997, MNRAS, 292, 887
\bibitem[\protect\citeauthoryear{Greene et al.}{2006}]{greene06} Greene, J. E., Ho, L. C.,  Ulvestad, J. S. 2006, ApJ, 636, 56
\bibitem[\protect\citeauthoryear{Hawley, Gammie \& Balbus}{1995}]{hawley95} Hawley, J. F., Gammie, C. F.,  Balbus, S. A. 1995, ApJ, 440, 742
\bibitem[\protect\citeauthoryear{Hawley \& Krolik}{2006}]{hawley06} Hawley, J. F.,  Krolik, J. H. 2006, ApJ, 641, 103
\bibitem[\protect\citeauthoryear{Hirose et al.}{2004}]{hirose04} Hirose, S., Krolik, J. H., De Villiers, J. P.,  Hawley, J. F. 2004, ApJ, 606, 1083
\bibitem[\protect\citeauthoryear{Igumenshchev et al.}{2003}]{igu03} Igumenshchev, I. V., Narayan, R.,  Abramowicz, M. A. 2003, ApJ, 592, 1042
\bibitem[\protect\citeauthoryear{Koide et al.}{2000}]{koide00} Koide, S., Meier, D. L., Shibata, K.,  Kudoh, T. 2000, ApJ, 536, 668
\bibitem[\protect\citeauthoryear{Koide}{2003}]{koide03} Koide, S. 2003, Phys. Rev. D, 67, 104010
\bibitem[\protect\citeauthoryear{Komissarov}{2005}]{komissarov05} Komissarov, S. S. 2005, MNRAS, 359, 801
\bibitem[\protect\citeauthoryear{K\"{o}rding, Jester \& Fender}{2006}]{kording06} K\"{o}rding, E. G., Jester, S.,  Fender, R. 2006, MNRAS, 372, 1366
\bibitem[\protect\citeauthoryear{Livio et al.}{1999}]{livio99} Livio, M., Ogilvie, G. I.,  Pringle, J. E. 1999, ApJ, 512, 100
\bibitem[\protect\citeauthoryear{Loewenstein et al.}{2001}]{loewen01} Loewenstein, M., Mushotzky, R. F., Angelini, L., Arnaud, K. A.,  Quataert, E. 2001, ApJ, 555, L21
\bibitem[\protect\citeauthoryear{Maccarone et al.}{2003}]{maccarone03} Maccarone, T. J., Gallo, E.,  Fender, R. 2003, MNRAS, 345, L19
\bibitem[\protect\citeauthoryear{Macdonald \& Thorne}{1982}]{macdonald82} Macdonald, D.,  Thorne, K. S. 1982, MNRAS, 198, 345
\bibitem[\protect\citeauthoryear{McCarthy et al.}{2004}]{mccarthy04} McCarthy, I. G., Balogh, M. L., Babul, A.,
Poole, G. B., Horner, D. J. 2004, ApJ, 613, 811
\bibitem[\protect\citeauthoryear{McKinney}{2005}]{mckinney05} McKinney, J. C. 2005, ApJ, 630, L5
\bibitem[\protect\citeauthoryear{McKinney \& Gammie}{2004}]{mckinney04} McKinney, J. C.,  Gammie, C. F. 2004, ApJ, 611, 977
\bibitem[\protect\citeauthoryear{Meier}{1999}]{meier99} Meier, D. L. 1999, ApJ, 522, 753
\bibitem[\protect\citeauthoryear{Meier}{2001}]{meier01} Meier, D. L. 2001, ApJ, 548, L9
\bibitem[\protect\citeauthoryear{Merloni et al.}{2003}]{merloni03} Merloni, A., Heinz, S.,  di Matteo, T. 2003, MNRAS, 345, 1057
\bibitem[\protect\citeauthoryear{Narayan, Mahadevan \& Quataert}{1998}]{narayan98} Narayan, R., Mahadevan, R.,  Quataert, E. 1998, in Abramowicz M. A., Bj\"ornsson G., Pringle J. E., eds, The Theory of Black Hole Accretion Disks.  Cambridge Univ. Press, p. 148 (N98)
\bibitem[\protect\citeauthoryear{Narayan}{2005}]{narayan05} Narayan, R. 2005, Ap\&SS, 300, 177
\bibitem[\protect\citeauthoryear{Narayan \& Yi}{1995}]{narayan95} Narayan, R.,  Yi, I. 1995, ApJ, 452, 710
\bibitem[\protect\citeauthoryear{Nemmen et al.}{2006}]{nemmen06} Nemmen, R. S., Storchi-Bergmann, T., Yuan, F.,
Eracleous, M., Terashima, Y., Wilson, A. S. 2006, ApJ, 643, 652
\bibitem[\protect\citeauthoryear{Novikov \& Thorne}{1973}]{novikov73} Novikov, I. D.,  Thorne, K. S. 1973, in Black Holes, De Witt C., De Witt B. S., eds, Black Holes. Gordon \& Breach, New York, p. 344
\bibitem[\protect\citeauthoryear{Nusser, Silk \& Babul}{2006}]{nusser06} Nusser, A., Silk, J.,  Babul, A. 2006, MNRAS, 373, 739
\bibitem[\protect\citeauthoryear{Pellegrini}{2005}]{p05} Pellegrini, S. 2005, ApJ, 624, 155
\bibitem[\protect\citeauthoryear{Proga \& Begelman}{2003}]{proga03} Proga, D.,  Begelman, M. C. 2003, ApJ, 592, 767
\bibitem[\protect\citeauthoryear{Punsly \& Coroniti}{1990}]{punsly90} Punsly, B.,  Coroniti, F. V. 1990, ApJ, 354, 583
\bibitem[\protect\citeauthoryear{Quataert \& Narayan}{1999}]{quataert99} Quataert, E.,  Narayan, R. 1999, ApJ, 520, 298
\bibitem[\protect\citeauthoryear{Quataert \& Gruzinov}{2000}]{quataert00} Quataert, E.,  Gruzinov, A.. 2000, ApJ, 539, 809
\bibitem[\protect\citeauthoryear{Rafferty et al.}{2006}]{rafferty06} Rafferty, D. A., McNamara, B. R., Nulsen, P. E. J.,  Wise, M. W. 2006, ApJ, 652, 216
\bibitem[\protect\citeauthoryear{Rees et al.}{1982}]{rees82} Rees, M. J., Begelman, M. C., Blandford, R. D.,  Phinney, E. S. 1982, Nature, 295, 17
\bibitem[\protect\citeauthoryear{Sikora, Stawarz \& Lasota}{2007}]{sikora07} Sikora, M., Stawarz, L.,  Lasota, J.-P. 2007, ApJ, 658, 815
\bibitem[\protect\citeauthoryear{Sijacki \& Springel}{2006}]{sijacki06} Sijacki, D.,  Springel, V. 2006, MNRAS, 366, 397
\bibitem[\protect\citeauthoryear{Shakura \& Sunyaev}{1973}]{shakura73} Shakura, N. I.,  Sunyaev, R. A. 1973, A\&A, 24, 337
\bibitem[\protect\citeauthoryear{Soltan}{1982}]{soltan82} Soltan, A. 1982, MNRAS, 200, 115
\bibitem[\protect\citeauthoryear{Soria et al.}{2006}]{soria06} Soria, R., Fabbiano, G., Graham, A. W., Baldi, A., Elvis,
M., Jerjen, H., Pellegrini, S., Siemiginowska, A. 2006, ApJ, 640, 126
\bibitem[\protect\citeauthoryear{Taylor et al.}{2006}]{taylor06} Taylor, G. B., Sanders, J. S., Fabian, A. C.,  Allen, S. W. 2006, MNRAS, 365, 705
\bibitem[\protect\citeauthoryear{Thorne}{1974}]{thorne74} Thorne, K. S. 1974, ApJ, 191, 507
\bibitem[\protect\citeauthoryear{Thorne et al.}{1986}]{tpm86} Thorne, K. S., Price, R. H.,  Macdonald, D. A., eds. 1986, Black Holes: The Membrane Paradigm. Yale Univ. Press, New Haven, p. 132
\bibitem[\protect\citeauthoryear{Tremaine et al.}{2002}]{tremaine02} Tremaine, S., et al. 2002, ApJ, 574, 740
\bibitem[\protect\citeauthoryear{Volonteri et al.}{2005}]{volonteri05} Volonteri, M., Madau, P., Quataert, E.,  Rees, M. 2005, ApJ, 620, 69
\bibitem[\protect\citeauthoryear{Wang et al.}{2006}]{wang06} Wang, J.-M., Chen, Y.-M., Ho, L. C.,  McLure, R. J. 2006, ApJ, 642, L111
\bibitem[\protect\citeauthoryear{Yu \& Tremaine}{2002}]{yu02} Yu, Q.,  Tremaine, S. 2002, MNRAS, 335, 965
\end{thebibliography}
\end{document}